\documentclass[aps,tightenlines]{revtex4}

\usepackage{amsfonts}
\usepackage{amsmath}
\usepackage{amssymb}
\usepackage{graphicx}

\begin{document}

\title{Generation of field mediated three qubit entangled state shared by
Alice and Bob }
\author{Paulo Jose dos Reis}
\affiliation{Departamento de F\'{\i}sica, Universidade Estadual de Londrina, Londrina
86051-990, PR Brazil }
\author{S. Shelly Sharma}
\email{shelly@uel.br}
\affiliation{Departamento de F\'{\i}sica, Universidade Estadual de Londrina, Londrina
86051-990, PR Brazil }

\begin{abstract}
A scheme to generate shared tripartite entangled states, with two-trapped
atoms in a cavity held by Alice (qubits $A_{1}$ and $A_{2}$) entangled to a
single trapped atom in a remote lab owned by Bob ($B$), is proposed. The
entanglement is generated through interaction of trapped atoms with two mode
squeezed light shared by the two cavities. The proposed scheme is an
extension of the proposal of ref. [W. Son, M. S. Kim, J. Lee, and D. Ahn, J.
Mod. Opt. 49, 1739 (2002)], where the possibility of entangling two remote
qubits using a bipartite continuous variable state was examined. While the
global negativity detects the free entanglement of the three atom mixed
state, the bound entanglement is detected by the negativity calculated from
pure state decomposition of the state operator. The partial negativities
calculated by selective partial transposition of the three atom mixed state
detect the pairwise entanglement of qubit pairs $A_{1}B$, $A_{2}B$, and $%
A_{1}A_{2}$. The entanglement of three atoms is found to be W-like, no GHZ
like quantum correlations being generated.
\end{abstract}

\maketitle

\section{ Introduction}

Quantum systems with discrete energy spectrum are often used for storing,
manipulating, and transmitting information encoded in quantum states
belonging to finite-dimensional Hilbert spaces. A two-level atom is an
example of two-dimensional quantum systems, universally known as qubits. All
possible superposition states of a two-level system can be expressed in
terms of linear combinations of two-dimensional orthonormal states defining
the standard computational basis. On the other hand, the electromagnetic
field is characterized by continuous variables that is observables with a
continuous spectrum of eigenvalues. Quantum information based on continuous
variables is encoded in quantum states belonging to infinite-dimensional
Hilbert spaces. The entanglement of remote qubits is a resource for
implementation of communication protocols \cite{eker91} and information
processing \cite{bouw00}\ using quantum systems. W. Son et al. \cite{son02}
have shown the possibility of entangling two remote qubits using a bipartite
continuous variable state. In this article, a scheme to generate shared
tripartite entangled states of two-trapped atoms in a cavity held by Alice (
qubits $A_{1}$ and $A_{2}$) and a single trapped atom in a remote lab owned
by Bob ($B$), is proposed. A single mode of the two-mode squeezed
electromagnetic field, generated by a squeezed light source, is injected
into each cavity by a beam splitter. The entanglement is generated through
interaction of trapped atoms with quantized electromagnetic field shared by
the two cavities. No direct interaction amongst the atoms takes place.

The dynamics of three atom entanglement due to atom field interaction is
investigated after tracing over the field degrees of freedom. We use the
global negativity and partial $K-$way negativities to analyze the
entanglement dynamics. Negativity \cite{zycz98}, based on Peres Horodecki 
\cite{pere96,horo96} criterion, has been shown to be an entanglement
monotone \cite{vida02}. A comparison of global negativity of three atom
mixed state with respect to qubit $B$ and the linear entropy calculated from
the state operator for qubit $B$ indicates that for certain ranges of
interaction parameter values the global negativity fails to detect the
entanglement amongst the atomic qubits. Using pure state decomposition (PSD)
of mixed three atom state, we recalculate the negativity by making a
weighted sum of pure states contributions to global negativity. The global
negativity calculated from pure state decomposition (PSDG) of the state
operator is not a proper measure of entanglement because it is not invariant
with respect to local operations and classical communication (LOCC). It
does, however, detect the bound entanglement being greater or equal to the
convex roof extension (CRE) of negativity shown to be an entanglement
monotone \cite{sooj03}. We use the global negativity and partial $K-$way
negativities of pure state decomposition, primarily, to understand how
quantum coherences present in the three atom state, evolve with interaction
time. The coherences of a multipartite composite system having $N$
subsystems can be quantified by partial $K-$way ($2\leq K\leq N$)
negativities \cite{shar06,shar081}. Partial $K-$way negativity is the
contribution of a $K-$way partial transpose to global negativity. In a three
qubit state, the entanglement arising due to $3-$way and $2-$way coherences 
\cite{shar07} is quantified by partial $3-$way and $2-$way negativities. For
canonical states, positive valued partial $3-$way and $2-$way negativities
measure, respectively, the three qubit GHZ-like and pairwise entanglement of
the system \cite{shar082}. We further use the partial negativities
calculated from the pure state decomposition of three atom mixed state to
study the dynamics of two-body coherences of qubit pairs $A_{1}B$, $A_{2}B$,
and $A_{1}A_{2}$. The qubits $A_{1}$ and $A_{2}$ are shown to form a bell
pair with certain probability, which is entangled to qubit $B$. The degree
of entanglement depends on the atom field coupling strength, squeeze
parameter, as well as the coefficient of reflection of the beam splitter.

\section{The cavity field}

Consider two cavities prepared in the vacuum state at $t=0$. The beam
splitter used to inject the external field into a cavity is represented by
the operator%
\begin{equation}
\hat{B}(\theta )=\exp \left[ \frac{\theta }{2}(\hat{c}\hat{f}^{\dagger }-%
\hat{c}^{\dagger }\hat{f})\right] ,  \label{1}
\end{equation}%
where $\hat{c}$ $\left( \hat{f}\right) $and $\hat{c}^{\dagger }\left( \hat{f}%
^{\dagger }\right) $are creation and annihilation operators for photons in
the cavity (external field), respectively. The coefficient $r=\cos \frac{%
\theta }{2}$ is the reflection coefficient of the beam splitter. The two
mode squeeze operator acting on two mode vacuum state

\begin{equation}
\hat{S}(s)\left\vert 0,0\right\rangle =\exp (-s\hat{a}\hat{b}+s\hat{a}%
^{\dagger }\hat{b}^{\dagger })\left\vert 0,0\right\rangle ,  \label{4}
\end{equation}%
produces a two-mode squeezed field in composite state

\begin{equation}
\left\vert \Psi _{F}\right\rangle =\frac{1}{\cosh s}\overset{\infty }{%
\underset{n=0}{\sum }}(\tanh s)^{n}\left\vert n,n\right\rangle ,  \label{5}
\end{equation}%
where $s$ is the squeeze parameter. The bosonic creation and annihilation
operators for mode one are $\hat{a}^{\dagger }$, $\hat{a}$ and for mode two
are $\hat{b}^{\dag }$, $\hat{b}$, respectively. The two mode squeezed state
is an entangled state having bipartite entanglement determined by the value
of squeeze parameter $s$. Using beam splitter (represented by operator of
Eq. (\ref{1})) to inject the modes one and two of the squeezed field (Eq. (%
\ref{5})) into independent cavities $c_{1}$ and $c_{2}$, respectively, the
cavity field state at $t=0$ reads as 
\begin{equation}
\hat{\rho}_{c_{1}c_{2}}(0)=\left( \frac{1}{\cosh s}\right) ^{2}\overset{%
\infty }{\underset{n,m=0}{\sum }}\underset{k,l=0}{\overset{\min \left[ n,m%
\right] }{\;\sum }}(\tanh s)^{n+m}G_{kl}^{nm}(\theta )\left\vert
n-k,n-l\right\rangle \left\langle m-k,m-l\right\vert ,  \label{6}
\end{equation}%
where 
\begin{equation}
G_{kl}^{nm}(\theta )=C_{k}^{n}\left( \theta \right) C_{k}^{m}\left( \theta
\right) C_{l}^{n}\left( \theta \right) C_{l}^{m}\left( \theta \right) ,
\label{7}
\end{equation}%
\begin{equation}
G_{kl}^{nn}(\theta )=\frac{n!}{k!(n-k)!}\frac{n!}{l!(n-l)!}\cos ^{2k}\left( 
\frac{\theta }{2}\right) \sin ^{2n-2k}\left( \frac{\theta }{2}\right) \cos
^{2l}\left( \frac{\theta }{2}\right) \sin ^{2n-2l}\left( \frac{\theta }{2}%
\right) ,  \label{8}
\end{equation}%
and 
\begin{equation}
C_{k}^{n}\left( \theta \right) =\sqrt{\frac{n!}{k!(n-k)!}}\cos ^{k}\left( 
\frac{\theta }{2}\right) \sin ^{n-k}\left( \frac{\theta }{2}\right) .
\label{9}
\end{equation}

The operator $\hat{\rho}_{c_{1}c_{2}}(0)$ represents a mixed composite state
of cavity field.

\section{Atom field interaction}

The interaction of N two-level atoms with a resonant single mode quantized
electromagnetic field in rotating wave approximation is described by Tavis
-Cummings model \cite{tavi68} Hamiltonian%
\begin{equation}
\hat{H}_{N}=\hbar \omega _{0}\widehat{a}^{\dagger }\widehat{a}+\frac{\hbar
\omega _{0}}{2}\sum\limits_{i=1,N}\widehat{\sigma }_{z}^{i}+\hbar
g\sum\limits_{i=1,N}\left[ \widehat{\sigma }_{+}^{i}\widehat{a}+\widehat{%
\sigma }_{-}^{i}\widehat{a}^{\dag }\right] ,  \label{10}
\end{equation}%
where $\omega _{0}$ is the atomic transition frequency and $g$ the atom
field coupling strength. Eigen states of Pauli operator $\hat{\sigma}%
_{z}^{i} $ model the internal states of the $i^{th}$ atom ($\mathit{i}=1,2$)
with eigenvalue $m_{\sigma }^{i}$ $=-1$ ($+1$) standing for the ground
(excited) state of the atom. Defining the composite operator, $\hat{\sigma}%
_{k}=\sum_{i}\hat{\sigma}_{k}^{i}$ where $k=(z,+,-)$, we may construct the
eigen basis of operators $\widehat{\sigma }^{2}$ and $\hat{\sigma}_{z}$ to
represent N atom internal states. A typical coupled basis vector is written
as $\left\vert \sigma ,m_{\sigma }\right\rangle $, with eigenvalues of $\hat{%
\sigma}^{2}$ and $\hat{\sigma}_{z}$ given by $\sigma (\sigma +2)$ and $%
m_{\sigma },$ respectively.

\begin{figure}[t]
\centering \includegraphics[width=3.75in,height=5.0in,angle=-90]{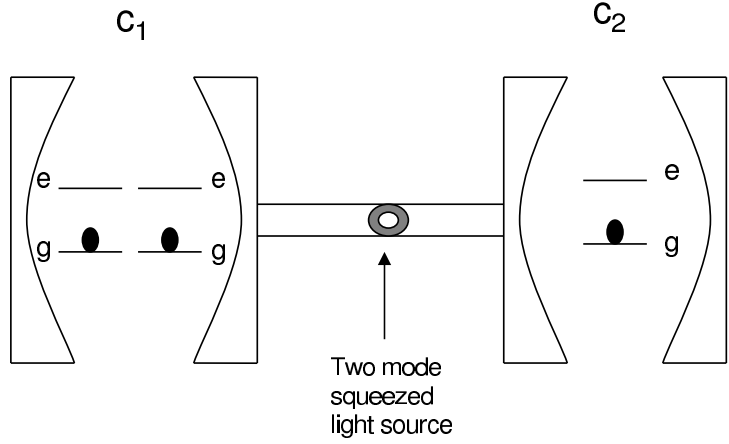}
\caption{The schematic diagram of experimental setup}
\label{fig0}
\end{figure}

Consider two atoms trapped in cavity $c_{1}$ radiated by the single mode
resonant cavity field in photon number state $\left\vert n\right\rangle $.
Interaction with the cavity field generates entanglement of internal states
of the atom, and the cavity field state. Working in the atom-field basis $%
\left\vert \sigma ,m_{\sigma },n\right\rangle $, the relevant basis states
for the case when two atoms are prepared in their ground states at $t=0$ are 
$\left\vert 2,-2,n\right\rangle $, $\left\vert 2,0,n-1\right\rangle $, and $%
\left\vert 2,2,n-2\right\rangle $ and the Hamiltonian matrix reads as%
\begin{equation}
H_{1I}^{n}=\left( 
\begin{array}{ccc}
0 & \hbar gB_{n} & 0 \\ 
\hbar gB_{n} & 0 & \hbar gA_{n} \\ 
0 & \hbar gA_{n} & 0%
\end{array}%
\right) ,  \label{12}
\end{equation}%
$\allowbreak $ where $A_{n}=\sqrt{2\left( n-1\right) }$, $B_{n}=\sqrt{2n}$.
Defining interaction parameter $\tau =gt,$ the unitary operator $\widehat{{\
U}}_{1}${\ $(\tau )$ }$=\exp \left[ -\left( i\widehat{H}_{1I}t\right)
/\hslash \right] $ is found to be 
\begin{equation}
\text{{\small $U_{1}^{n}(\tau )=\left( 
\begin{array}{ccc}
\frac{\left[ B_{n}^{2}\cos \left( f_{n}\tau \right) +A_{n}^{2}\right] }{%
A_{n}^{2}+B_{n}^{2}} & \frac{-iB_{n}\sin \left( f_{n}\tau \right) }{\sqrt{%
\left( A_{n}^{2}+B_{n}^{2}\right) }} & \frac{A_{n}B_{n}\left[ \cos \left(
f_{n}\tau \right) -1\right] }{A_{n}^{2}+B_{n}^{2}} \\ 
\frac{-iB_{n}\sin \left( f_{n}\tau \right) }{\sqrt{\left(
A_{n}^{2}+B_{n}^{2}\right) }} & \cos \left( f_{n}\tau \right) & \frac{%
-iA_{n}\sin \left( f_{n}\tau \right) }{\sqrt{\left(
A_{n}^{2}+B_{n}^{2}\right) }} \\ 
\frac{A_{n}B_{n}\left[ \cos \left( f_{n}\tau \right) -1\right] }{%
A_{n}^{2}+B_{n}^{2}} & -iA_{n}\sin \left( f_{n}g\tau \right) & \frac{\left[
A_{n}^{2}\cos \left( f_{n}\tau \right) +B_{n}^{2}\right] }{%
A_{n}^{2}+B_{n}^{2}}%
\end{array}%
\right) ,$}}  \label{13}
\end{equation}%
where $f_{n}=\sqrt{2\left( 2n-1\right) }.$

For a single atom interacting with the cavity field in cavity $c_{2}$, the
unitary matrix that determines the time evolution of the system in the basis 
$\left\vert 1,-1,m\right\rangle ,\left\vert 1,1,m-1\right\rangle $ is%
\begin{equation}
U_{2}^{m}(\tau )=\allowbreak \left( 
\begin{array}{cc}
\cos \left( \sqrt{m}\tau \right) & -i\sin \left( \sqrt{m}\tau \right) \\ 
i\sin \left( \sqrt{m}\tau \right) & \cos \left( \sqrt{m}\tau \right)%
\end{array}%
\right) \text{.}  \label{11}
\end{equation}

The evolution operator $\widehat{U}_{12}^{nm}(\tau )$ for the two cavity
composite system is obtained by taking the tensor product 
\begin{equation}
\widehat{U}_{12}^{nm}(\tau )=\widehat{U}_{1}^{n}(\tau )\otimes \widehat{U}%
_{2}^{m}(\tau ).  \label{14}
\end{equation}

\section{Three qubit State operator}

Consider two-trapped atoms in cavity $c_{1}$ held by Alice ( qubits $A_{1}$
and $A_{2}$) and a single trapped atom (qubit $B$) in a remote cavity $c_{2}$
controlled by Bob. The atoms are cooled down to their ground state with
cavities prepared in vacuum state at $t=0$. A squeezed light source
generates the two mode electromagnetic field. A schematic diagram of the
setup is shown in figure(\ref{fig0}). The two mode squeezed field is
injected into the cavities, with each cavity receiving a single mode of the
field. The composite state of field in two cavities is given by $\hat{\rho}%
_{c_{1}c_{2}}(0)$ as in Eq. (\ref{6}). The state of the composite system
without atom field interaction is represented by 
\begin{equation}
\hat{\rho}_{(0)}=\widehat{\rho }^{A_{1}A_{2}B}(0)\otimes \hat{\rho}%
_{c_{1}c_{2}}(0),  \label{3}
\end{equation}%
where $\widehat{\rho }^{A_{1}A_{2}B}(0)=\left\vert 2,-2\right\rangle
_{c_{1}}\left\vert 1,-1\right\rangle _{c_{2}}\left\langle 1,-1\right\vert
_{c_{1}}\left\langle 2,-2\right\vert _{c_{2}}$ in the coupled basis (basis
states labelled by eigenvalues of $\hat{\sigma}^{2}$ and $\hat{\sigma}_{z}).$
Here $\left\vert 2,-2\right\rangle _{c_{1}}$ represents qubits $A_{1}$ and $%
A_{2}$ in state $\left\vert 2,-2\right\rangle $ in cavity $c_{1}$. The state
operator, after interaction between the atoms and the respective squeezed
fields in the two cavities for a time $\ t=\tau /g$, is given by 
\begin{equation}
\widehat{\rho }(\tau )=\widehat{U}_{12}(\tau )\widehat{\rho }%
^{A_{1}A_{2}B}(0)\otimes \hat{\rho}_{c_{1}c_{2}}(0)\widehat{U}_{12}^{\dagger
}(\tau )  \label{3a}
\end{equation}%
Using the unitary evolution operators due to atom-field interaction in
cavities one and two (Eq. (\ref{14})), the state of the composite system
after interaction time $t$ is written as

\begin{eqnarray}
\hat{\rho}(\tau ) &=&\left( \frac{1}{\cosh s}\right) ^{2}\hat{U}_{12}(\tau )%
\left[ \overset{\infty }{\underset{n,m=0}{\sum }}(\tanh s)^{n+m}\right. 
\notag \\
&&\left. \underset{k,l=0}{\overset{\min \left[ n,m\right] }{\;\sum }}%
G_{kl}^{nm}(\theta )\left\vert 2,-2,n-k\right\rangle \left\vert
1,-1,n-l\right\rangle \left\langle 2,-2,m-k\right\vert \left\langle
1,-1,m-l\right\vert \right] \hat{U}_{12}^{\dagger }(\tau )  \notag \\
&=&\left( \frac{1}{\cosh s}\right) ^{2}\overset{\infty }{\underset{n,m=0}{%
\sum }}\underset{k,l=0}{\overset{\min \left[ n,m\right] }{\;\sum }}(\tanh
s)^{n+m}G_{kl}^{nm}(\theta )\left\{ \hat{U}_{1}^{n-k}(\tau )\otimes \hat{U}%
_{2}^{n-l}(\tau )\right.  \notag \\
&&\left. \left\vert 2,-2,n-k\right\rangle \left\vert 1,-1,n-l\right\rangle
\left\langle 2,-2,m-k\right\vert \left\langle 1,-1,m-l\right\vert \hat{U}%
_{1}^{\dagger m-k}(\tau )\otimes \hat{U}_{2}^{\dagger m-l}(\tau )\right\} .
\end{eqnarray}%
The reduced state operator for atoms is obtained from $\widehat{\rho }(\tau
),$ by tracing over the field modes, that is%
\begin{equation}
\widehat{\rho }^{A_{1}A_{2}B}(\tau )=Tr_{c_{1}c_{2}}\left( \hat{\rho}(\tau
)\right) .
\end{equation}

Associating, single qubit states $\left\vert 0\right\rangle $ ($\left\vert
1\right\rangle )$ with ground (excited) state of the two-level atom, the
initial state of three atoms in computational basis reads as $\widehat{\rho }%
^{A_{1}A_{2}B}(0)=\left( \left\vert 00\right\rangle _{c_{1}}\left\langle
00\right\vert \right) \otimes \left( \left\vert 0\right\rangle
_{c_{2}}\left\langle 0\right\vert \right) $. The matrix $\hat{\rho}%
^{A_{1}A_{2}B}(\tau )$ in the computational basis $\left\vert
00\right\rangle _{c_{1}}\left\vert 0\right\rangle _{c_{2}}$, $\left\vert
10\right\rangle _{c_{1}}\left\vert 0\right\rangle _{c_{2}}$, $\left\vert
01\right\rangle _{c_{1}}\left\vert 0\right\rangle _{c_{2}}$, $\left\vert
11\right\rangle _{c_{1}}\left\vert 0\right\rangle _{c_{2}}$, $\left\vert
00\right\rangle _{c_{1}}\left\vert 1\right\rangle _{c_{2}}$, $\left\vert
10\right\rangle _{c_{1}}\left\vert 1\right\rangle _{c_{2}}$, $\left\vert
01\right\rangle _{c_{1}}\left\vert 1\right\rangle _{c_{2}}$, $\left\vert
11\right\rangle _{c_{1}}\left\vert 1\right\rangle _{c_{2}}$, reads as

\begin{equation}
\widehat{\rho }^{A_{1}A_{2}B}(\tau )=\left( 
\begin{array}{cccccccc}
\rho _{11}^{A_{1}A_{2}B} & 0 & 0 & 0 & 0 & \frac{\rho _{15}^{A_{1}A_{2}B}}{%
\sqrt{2}} & \frac{\rho _{15}^{A_{1}A_{2}B}}{\sqrt{2}} & 0 \\ 
0 & \frac{\rho _{22}^{A_{1}A_{2}B}}{2} & \frac{\rho _{22}^{A_{1}A_{2}B}}{2}
& 0 & 0 & 0 & 0 & \frac{\rho _{26}^{A_{1}A_{2}B}}{\sqrt{2}} \\ 
0 & \frac{\rho _{22}^{A_{1}A_{2}B}}{2} & \frac{\rho _{22}^{A_{1}A_{2}B}}{2}
& 0 & 0 & 0 & 0 & \frac{\rho _{26}^{A_{1}A_{2}B}}{\sqrt{2}} \\ 
0 & 0 & 0 & \rho _{33}^{A_{1}A_{2}B} & 0 & 0 & 0 & 0 \\ 
0 & 0 & 0 & 0 & \rho _{44}^{A_{1}A_{2}B} & 0 & 0 & 0 \\ 
\frac{\rho _{15}^{A_{1}A_{2}B}}{\sqrt{2}} & 0 & 0 & 0 & 0 & \frac{\rho
_{55}^{A_{1}A_{2}B}}{2} & \frac{\rho _{55}^{A_{1}A_{2}B}}{2} & 0 \\ 
\frac{\rho _{15}^{A_{1}A_{2}B}}{\sqrt{2}} & 0 & 0 & 0 & 0 & \frac{\rho
_{55}^{A_{1}A_{2}B}}{2} & \frac{\rho _{55}^{A_{1}A_{2}B}}{2} & 0 \\ 
0 & \frac{\rho _{26}^{A_{1}A_{2}B}}{\sqrt{2}} & \frac{\rho
_{26}^{A_{1}A_{2}B}}{\sqrt{2}} & 0 & 0 & 0 & 0 & \rho _{66}^{A_{1}A_{2}B}%
\end{array}%
\right) \text{,}  \label{roatomt}
\end{equation}%
with the matrix elements given by 
\begin{eqnarray}
\rho _{11}^{A_{1}A_{2}B}(\tau ) &=&\left( \frac{1}{\cosh s}\right) ^{2}%
\overset{\infty }{\underset{n=0}{\sum }}\underset{k,l=0}{\overset{n}{\;\sum }%
}(\tanh s)^{2n}G_{kl}^{nn}(\theta )\cos ^{2}\left( \tau \sqrt{n-l}\right) 
\notag \\
&&\left( \frac{n-k-1}{2n-2k-1}+\frac{n-k}{2n-2k-1}\cos \left( f_{nk}\tau
\right) \right) ^{2},
\end{eqnarray}

\begin{eqnarray}
\rho _{22}^{A_{1}A_{2}B}(\tau ) &=&\left( \frac{1}{\cosh s}\right) ^{2}%
\overset{\infty }{\underset{n=0}{\sum }}\underset{k,l=0}{\overset{n}{\;\sum }%
}(\tanh s)^{2n}G_{kl}^{nn}(\theta )\cos ^{2}\left( \tau \sqrt{n-l}\right) 
\notag \\
&&\frac{\left( n-k\right) }{(2n-2k-1)}\sin ^{2}\left( f_{nk}\tau \right) ,
\end{eqnarray}%
\begin{eqnarray}
\rho _{33}^{A_{1}A_{2}B}(\tau ) &=&\left( \frac{1}{\cosh s}\right) ^{2}%
\overset{\infty }{\underset{n=0}{\sum }}\underset{k,l=0}{\overset{n}{\;\sum }%
}(\tanh s)^{2n}G_{kl}^{nn}(\theta )\cos ^{2}\left( \tau \sqrt{n-l}\right) 
\notag \\
&&\frac{\left( n-k\right) \left( n-k-1\right) }{(2n-2k-1)^{2}}\left( \cos
\left( f_{nk}\tau \right) -1\right) ^{2},
\end{eqnarray}

\begin{eqnarray}
\rho _{44}^{A_{1}A_{2}B}(\tau ) &=&\left( \frac{1}{\cosh s}\right) ^{2}%
\overset{\infty }{\underset{n=0}{\sum }}\underset{k,l=0}{\overset{n}{\;\sum }%
}(\tanh s)^{2n}G_{kl}^{nn}(\theta )\sin ^{2}\left( \tau \sqrt{n-l}\right) 
\notag \\
&&\frac{1}{(2n-2k-1)^{2}}\left( \left( n-k-1\right) +\left( n-k\right) \cos
\left( f_{nk}\tau \right) \right) ^{2},
\end{eqnarray}

\begin{eqnarray}
\rho _{55}^{A_{1}A_{2}B}(\tau ) &=&\left( \frac{1}{\cosh s}\right) ^{2}%
\overset{\infty }{\underset{n=0}{\sum }}\underset{k,l=0}{\overset{n}{\;\sum }%
}(\tanh s)^{2n}G_{kl}^{nn}(\theta )\sin ^{2}\left( \tau \sqrt{n-l}\right) 
\notag \\
&&\frac{\left( n-k\right) }{(2n-2k-1)}\sin ^{2}\left( f_{nk}\tau \right) ,
\end{eqnarray}

\begin{eqnarray}
\rho _{66}^{A_{1}A_{2}B}(\tau ) &=&\left( \frac{1}{\cosh s}\right) ^{2}%
\overset{\infty }{\underset{n=0}{\sum }}\underset{k,l=0}{\overset{n}{\;\sum }%
}(\tanh s)^{2n}G_{kl}^{nn}(\theta )\sin ^{2}\left( \tau \sqrt{n-l}\right) 
\notag \\
&&\frac{\left( n-k\right) \left( n-k-1\right) }{(2n-2k-1)^{2}}\left( \cos
\left( f_{nk}\tau \right) -1\right) ^{2},
\end{eqnarray}%
\begin{eqnarray}
\rho _{51}^{A_{1}A_{2}B}(\tau ) &=&\left( \frac{1}{\cosh s}\right) ^{2}%
\overset{\infty }{\underset{n=0}{\sum }}\underset{k,l=0}{\overset{n}{\;\sum }%
}(\tanh s)^{2n+1}G_{kl}^{nn+1}(\theta )\frac{\sqrt{\left( n-k+1\right) }}{%
\sqrt{(2n-2k+1)}}  \notag \\
&&\cos \left( \tau \sqrt{n-l}\right) \sin \left( \tau \sqrt{n-l+1}\right)
\sin \left( f_{n+1k}\tau \right)  \notag \\
&&\left( \frac{\left( n-k-1\right) }{(2n-2k-1)}+\frac{\left( n-k\right) }{%
(2n-2k-1)}\cos \left( f_{nk}\tau \right) \right) ,
\end{eqnarray}%
\begin{eqnarray}
\rho _{62}^{A_{1}A_{2}B}(\tau ) &=&\left( \frac{1}{\cosh s}\right) ^{2}%
\overset{\infty }{\underset{n=0}{\sum }}\underset{k,l=0}{\overset{n}{\;\sum }%
}(\tanh s)^{2n+1}G_{kl}^{nn+1}(\theta )\frac{\sqrt{\left( n-k-1\right) }}{%
\sqrt{(2n-2k-1)}}  \notag \\
&&\frac{\sqrt{(n-k)\left( n-k+1\right) }}{(2n-2k+1)}\cos \left( \tau \sqrt{%
n-l}\right)  \notag \\
&&\sin \left( \tau \sqrt{n-l+1}\right) \left( \sin \left( f_{nk}\tau \right)
\right) \left( \cos \left( f_{n+1k}\tau \right) -1\right) ,
\end{eqnarray}%
and $f_{nk}=\sqrt{2(2n-2k-1)}.$

\begin{figure}[t]
\centering \includegraphics[width=3.75in,height=5.0in,angle=-90]{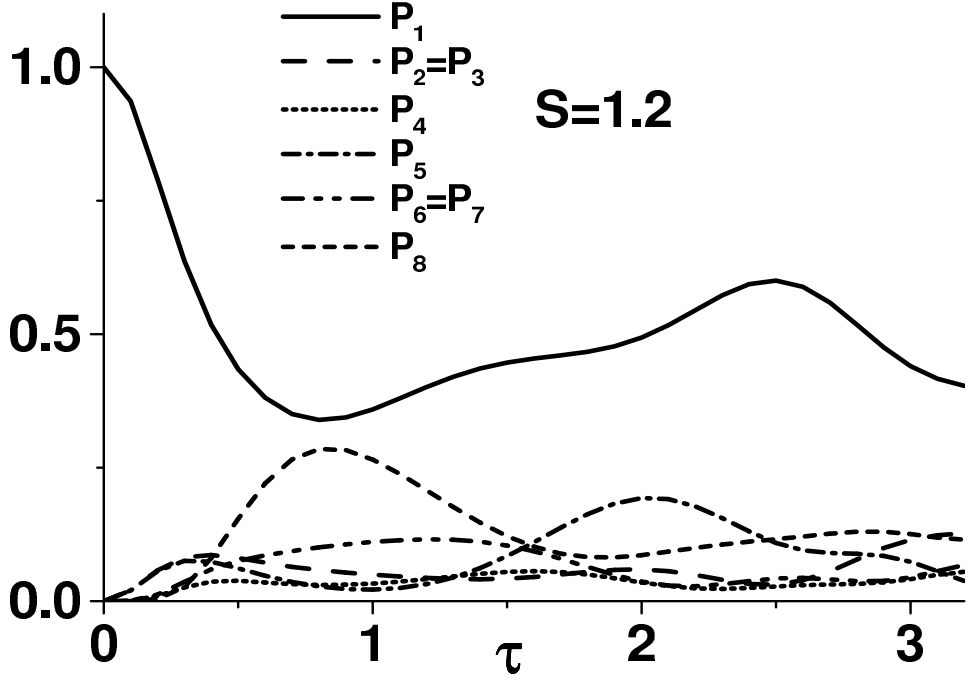}
\caption{The probabilities, versus $\protect\tau (=g\protect\eta t)$ for $%
s=1.2$.}
\label{fig1}
\end{figure}

\section{Probability of generating three qubit entangled states}

We have calculated, numerically, the matrix elements of state operator $%
\widehat{\rho }^{A_{1}A_{2}B}(\tau )$ using a simple fortran program. The
sum $\overset{\infty }{\underset{n=0}{\sum }}$ in Eq. (\ref{6}) is
approximated by $\overset{n_{\max }}{\underset{n=0}{\sum }}$ , where the
value of $n_{\max }$ is determined by the choice of parameter $s$. By
analyzing the variation of $\rho _{ii}^{A_{1}A_{2}B}(\tau )$ ($i=1-8$) with
parameter $s$ we have chosen $s=1.2$ as the preferred value of squeeze
parameter and $n_{\max }=80$. A direct observation of $-10$ dB squeezing of
quantum noise of light, reported recently by H. Vahlbruch et al. \cite%
{vahl08}, makes the selected $s$ value a reasonable choice.

Figure (\ref{fig1}) displays the probabilities $P_{i}$ ($i=1-8$) of finding
the three atoms in the local basis states $\left\vert
i_{1}i_{2}\right\rangle _{c_{1}}\left\vert i_{3}\right\rangle _{c_{2}}$ ($%
i=i_{1}+2i_{2}+4i_{3}+1),$ for $s=1.2$ and $\theta =\pi $ . We notice that
the probability of finding the atoms in their ground states drops to $0.34$
at $\tau =0.8,$ which is already less than $\sum\limits_{i=2-8}P_{i}$ at $%
\tau \sim 0.4.$ For $s=1.2$, the probability of initial atomic state
interacting with the field is reasonably large, while the mean photon number
is not too large. In case the mean photon number is large, the cavity decay
must be taken into account.

The operator $\widehat{\rho }^{A_{1}A_{2}B}(\tau )$ of Eq. (\ref{roatomt}),
can be rewritten in diagonal form as%
\begin{equation}
\widehat{\rho }^{A_{1}A_{2}B}(\tau )=\sum_{i=1}^{6}p_{i}\left\vert \Phi
_{i}\right\rangle \left\langle \Phi _{i}\right\vert \ ,  \label{ropured}
\end{equation}%
where $p_{i}$, and $\left\vert \Phi _{i}\right\rangle $ are the eigenvalues
and eigenvectors of $\widehat{\rho }^{A_{1}A_{2}B}(\tau )$ given by%
\begin{eqnarray}
\left\vert \Phi _{1}\right\rangle &=&a_{000}\left\vert 00\right\rangle
_{c_{1}}\left\vert 0\right\rangle _{c_{2}}+a_{101}\left( \frac{\left\vert
10\right\rangle _{c_{1}}+\left\vert 01\right\rangle _{c_{1}}}{\sqrt{2}}%
\right) \left\vert 1\right\rangle _{c_{2}},  \notag \\
p_{1} &=&\frac{\left( \rho _{11}^{A}+\rho _{55}^{A}\right) }{2}-\frac{1}{2}%
\sqrt{\left( \rho _{11}^{A}-\rho _{55}^{A}\right) ^{2}+4\left( \rho
_{15}^{A}\right) ^{2}},
\end{eqnarray}%
\begin{align}
\left\vert \Phi _{2}\right\rangle & =b_{000}\left\vert 00\right\rangle
_{c_{1}}\left\vert 0\right\rangle _{c_{2}}+b_{101}\left( \frac{\left\vert
10\right\rangle _{c_{1}}+\left\vert 01\right\rangle _{c_{1}}}{\sqrt{2}}%
\right) \left\vert 1\right\rangle _{c_{2}},  \notag \\
\qquad p_{2}& =\frac{\left( \rho _{11}^{A}+\rho _{55}^{A}\right) }{2}+\frac{1%
}{2}\sqrt{\left( \rho _{11}^{A}-\rho _{55}^{A}\right) ^{2}+4\left( \rho
_{15}^{A}\right) ^{2}},
\end{align}%
\begin{align}
\left\vert \Phi _{3}\right\rangle & =c_{100}\left( \frac{\left\vert
10\right\rangle _{c_{1}}+\left\vert 01\right\rangle _{c_{1}}}{\sqrt{2}}%
\right) \left\vert 0\right\rangle _{c_{2}}+c_{111}\left\vert 11\right\rangle
_{c_{1}}\left\vert 1\right\rangle _{c_{2}},  \notag \\
\qquad p_{3}& =\frac{\rho _{22}^{A}+\rho _{66}^{A}}{2}+\frac{1}{2}\sqrt{%
\left( \rho _{22}^{A}-\rho _{66}^{A}\right) ^{2}+4\rho _{26}^{2}},
\end{align}%
\begin{align}
\left\vert \Phi _{4}\right\rangle & =d_{100}\left( \frac{\left\vert
10\right\rangle _{c_{1}}+\left\vert 01\right\rangle _{c_{1}}}{\sqrt{2}}%
\right) \left\vert 0\right\rangle _{c_{2}}+d_{111}\left\vert 11\right\rangle
_{c_{1}}\left\vert 1\right\rangle _{c_{2}},  \notag \\
\qquad p_{4}& =\frac{\rho _{22}^{A}+\rho _{66}^{A}}{2}-\frac{1}{2}\sqrt{%
\left( \rho _{22}^{A}-\rho _{66}^{A}\right) ^{2}+4\rho _{26}^{2}},
\end{align}%
\begin{align}
\left\vert \Phi _{5}\right\rangle & =\left\vert 11\right\rangle
_{c_{1}}\left\vert 0\right\rangle _{c_{2}},\qquad p_{5}=\rho _{33}^{A}, 
\notag \\
\left\vert \Phi _{6}\right\rangle & =\left\vert 00\right\rangle
_{c_{1}}\left\vert 1\right\rangle _{c_{2}},\qquad p_{6}=\rho _{44}^{A}.
\end{align}%
The coefficients $a_{000}$, $a_{101}$, $b_{000}$, $b_{101}$, $c_{100}$, $%
c_{111}$, and $d_{100}$, $d_{111}$ are known in terms of the matrix elements
of $\widehat{\rho }^{A_{1}A_{2}B}(\tau ).$ The probabilities $p_{i}$
determine the entanglement distribution amongst the three qubits in the
system. The states $\left\vert \Phi _{i}\right\rangle ,$ $i=1\ $to $4$ are
W-like states. The state $\widehat{\rho }^{A_{1}A_{2}B}(\tau )$ \ is a
mixture of W-like states. For example, the fidelity of projecting out the
state 
\begin{equation}
\left\vert W_{1}\right\rangle =\frac{1}{\sqrt{3}}\left( \left\vert
00\right\rangle _{c_{1}}\left\vert 0\right\rangle _{c_{2}}+\left\vert
10\right\rangle _{c_{1}}\left\vert 1\right\rangle _{c_{2}}+\left\vert
01\right\rangle _{c_{1}}\left\vert 1\right\rangle _{c_{2}}\right) ,
\end{equation}%
is given by 
\begin{equation}
F(W_{1})=Tr\left( \widehat{\rho }^{A_{1}A_{2}B}(\tau )\left\vert
W_{1}\right\rangle \left\langle W_{1}\right\vert \right) .  \label{fidelity}
\end{equation}%
Alice can perform a controlled not gate on qubits $A_{1}$ and $A_{2}$
transforming the state into a mixture of GHZ like states.

It is easily verified that the probability of projecting out the qubits $%
A_{1}A_{2}$ in state $\left( \frac{\left\vert 10\right\rangle
_{c_{1}}+\left\vert 01\right\rangle _{c_{1}}}{\sqrt{2}}\right) $ from the
reduced state $\widehat{\rho }^{A_{1}A_{2}}(\tau )=Tr_{B}\left( \widehat{%
\rho }^{A_{1}A_{2}B}(\tau )\right) $ is $\left( \rho _{11}^{A}+\rho
_{55}^{A}\right) .$

\section{Three qubit mixed state entanglement dynamics}

The global partial transpose \cite{vida02} of $\widehat{\rho }%
^{A_{1}A_{2}B}(\tau )$ with respect to sub-systems, $A_{1}$, $A_{2}$, or $B$
is obtained by using the relations 
\begin{align}
\left\langle i_{1}i_{2}i_{3}\right\vert \rho _{G}^{T_{A_{1}}}(\tau
)\left\vert j_{1}j_{2}j_{3}\right\rangle & =\left\langle
j_{1}i_{2}i_{3}\right\vert \widehat{\rho }^{A_{1}A_{2}B}(\tau )\left\vert
i_{1}j_{2}j_{3}\right\rangle ,  \notag \\
\left\langle i_{1}i_{2}i_{3}\right\vert \rho _{G}^{T_{A_{2}}}(\tau
)\left\vert j_{1}j_{2}j_{3}\right\rangle & =\left\langle
i_{1}j_{2}i_{3}\right\vert \widehat{\rho }^{A_{1}A_{2}B}(\tau )\left\vert
j_{1}i_{2}j_{3}\right\rangle ,  \notag \\
\left\langle i_{1}i_{2}i_{3}\right\vert \rho _{G}^{T_{B}}(\tau )\left\vert
j_{1}j_{2}j_{3}\right\rangle & =\left\langle i_{1}i_{2}j_{3}\right\vert 
\widehat{\rho }^{A_{1}A_{2}B}(\tau )\left\vert j_{1}j_{2}i_{3}\right\rangle 
\text{.}
\end{align}%
The\ $K-$way partial transpose \cite{shar082} ($K$ $=2,3$) of tripartite
state $\widehat{\rho }^{A_{1}A_{2}B}$ with respect to qubits $A_{1}$, $A_{2}$%
, or $B$ is constructed by applying the following constraints: 
\begin{align}
\left\langle i_{1}i_{2}i_{3}\right\vert \widehat{\rho }_{K}^{T_{A_{1}}}\left%
\vert j_{1}j_{2}j_{3}\right\rangle & =\left\langle
j_{1}i_{2}i_{3}\right\vert \widehat{\rho }^{A_{1}A_{2}B}\left\vert
i_{1}j_{2}j_{3}\right\rangle ,\quad \text{if}\quad
\sum\limits_{m=1}^{3}(1-\delta _{i_{m},j_{m}})=K,  \notag \\
\left\langle i_{1}i_{2}i_{3}\right\vert \widehat{\rho }_{K}^{T_{A_{1}}}\left%
\vert j_{1}j_{2}j_{3}\right\rangle & =\left\langle
i_{1}i_{2}i_{3}\right\vert \widehat{\rho }^{A_{1}A_{2}B}\left\vert
j_{1}j_{2}j_{3}\right\rangle \quad \text{if}\quad
\sum\limits_{m=1}^{3}(1-\delta _{i_{m},j_{m}})\neq K,
\end{align}%
\begin{align}
\left\langle i_{1}i_{2}i_{3}\right\vert \widehat{\rho }_{K}^{T_{A_{2}}}\left%
\vert j_{1}j_{2}j_{3}\right\rangle & =\left\langle
i_{1}j_{2}i_{3}\right\vert \widehat{\rho }^{A_{1}A_{2}B}\left\vert
j_{1}i_{2}j_{3}\right\rangle ,\quad \text{if}\quad
\sum\limits_{m=1}^{3}(1-\delta _{i_{m},j_{m}})=K,  \notag \\
\left\langle i_{1}i_{2}i_{3}\right\vert \widehat{\rho }_{K}^{T_{A_{2}}}\left%
\vert j_{1}j_{2}j_{3}\right\rangle & =\left\langle
i_{1}i_{2}i_{3}\right\vert \widehat{\rho }^{A_{1}A_{2}B}\left\vert
j_{1}j_{2}j_{3}\right\rangle \quad \text{if}\quad
\sum\limits_{m=1}^{3}(1-\delta _{i_{m},j_{m}})\neq K,
\end{align}%
and%
\begin{align}
\left\langle i_{1}i_{2}i_{3}\right\vert \widehat{\rho }_{K}^{T_{B}}\left%
\vert j_{1}j_{2}j_{3}\right\rangle & =\left\langle
i_{1}i_{2}j_{3}\right\vert \widehat{\rho }^{A_{1}A_{2}B}\left\vert
j_{1}j_{2}i_{3}\right\rangle ,\quad \text{if}\quad
\sum\limits_{m=1}^{3}(1-\delta _{i_{m},j_{m}})=K,  \notag \\
\left\langle i_{1}i_{2}i_{3}\right\vert \widehat{\rho }_{K}^{T_{B}}\left%
\vert j_{1}j_{2}j_{3}\right\rangle & =\left\langle
i_{1}i_{2}i_{3}\right\vert \widehat{\rho }^{A_{1}A_{2}B}\left\vert
j_{1}j_{2}j_{3}\right\rangle \quad \text{if}\quad
\sum\limits_{m=1}^{3}(1-\delta _{i_{m},j_{m}})\neq K,
\end{align}%
where $\delta _{i_{m},j_{m}}=1$ for $i_{m}=j_{m}$, and $\delta
_{i_{m},j_{m}}=0$ for $i_{m}\neq j_{m}$.

The negative part of $\ \widehat{\rho }_{G}^{T_{p}}$ ($p$ $=A_{1}$ or $A_{2}$
or $B$) contains information about the amount and nature of entanglement
present in the tripartite system. By using a $K-$way partial transpose one
can extract the quantity of $K-$way coherence (that is the quantum
correlations present in a $K-$qubit GHZ like state) in the three qubit
state. The contribution of a $K-$way partial transpose to the global
negativity is calculated by noting that the global partial transpose with
respect to qubit $p$, may be written as 
\begin{equation}
\widehat{\rho }_{G}^{T_{p}}=\widehat{\rho }_{3}^{T_{p}}+\widehat{\rho }%
_{2}^{T_{p}}-\widehat{\rho }.  \label{3n}
\end{equation}%
Using $Tr\left( \widehat{\rho }_{G}^{T_{p}}\right) =1,$ the negativity of $%
\widehat{\rho }_{G}^{T_{p}}$ is given by 
\begin{equation}
{N}_{G}^{p}=-2\sum\limits_{i}\left\langle \Psi _{i}^{G-}\right\vert \widehat{%
\rho }_{G}^{T_{p}}\left\vert \Psi _{i}^{G-}\right\rangle
=-2\sum\limits_{i}\lambda _{i}^{G-}\text{,}  \label{2n}
\end{equation}%
where $\lambda _{i}^{G-}$and $\left\vert \Psi _{i}^{G-}\right\rangle $ are,
respectively, the $i^{th}$ negative eigenvalue and eigenvector of $\widehat{%
\rho }_{G}^{T_{p}}$. Substituting Eq. (\ref{3n}) in Eq. (\ref{2n}), we get%
\begin{equation}
-2\sum\limits_{i}\lambda
_{i}^{G-}=-2\sum\limits_{K=2}^{N}\sum\limits_{i}\left\langle \Psi
_{i}^{G-}\right\vert \widehat{\rho }_{K}^{T_{p}}\left\vert \Psi
_{i}^{G-}\right\rangle +2\sum\limits_{i}\left\langle \Psi
_{i}^{G-}\right\vert \widehat{\rho }\left\vert \Psi _{i}^{G-}\right\rangle .
\end{equation}%
Defining partial $K-$way negativity $E_{K}^{p}$ ($K=2$ to $N$) as 
\begin{equation}
E_{K}^{p}=-2\sum\limits_{i}\left\langle \Psi _{i}^{G-}\right\vert \widehat{%
\rho }_{K}^{T_{p}}\left\vert \Psi _{i}^{G-}\right\rangle ,\qquad \text{while 
}\qquad E_{0}^{p}=-2\sum\limits_{i}\left\langle \Psi _{i}^{G-}\right\vert 
\widehat{\rho }\left\vert \Psi _{i}^{G-}\right\rangle ,  \label{4n}
\end{equation}%
we may split the global negativity for qubit $p$ as%
\begin{equation}
N_{G}^{p}=E_{3}^{p}+E_{2}^{p}-E_{0}^{p}.  \label{5n}
\end{equation}%
The necessary condition for an $N-$partite pure state not to have $N-$%
partite entanglement is that the global negativity is zero for at least one
of the subsystems that is $N_{G}^{p}=0,$ where $p$ is one part in a
bipartite split of the composite system.

\begin{figure}[t]
\centering \includegraphics[width=3.75in,height=5.0in,angle=-90]{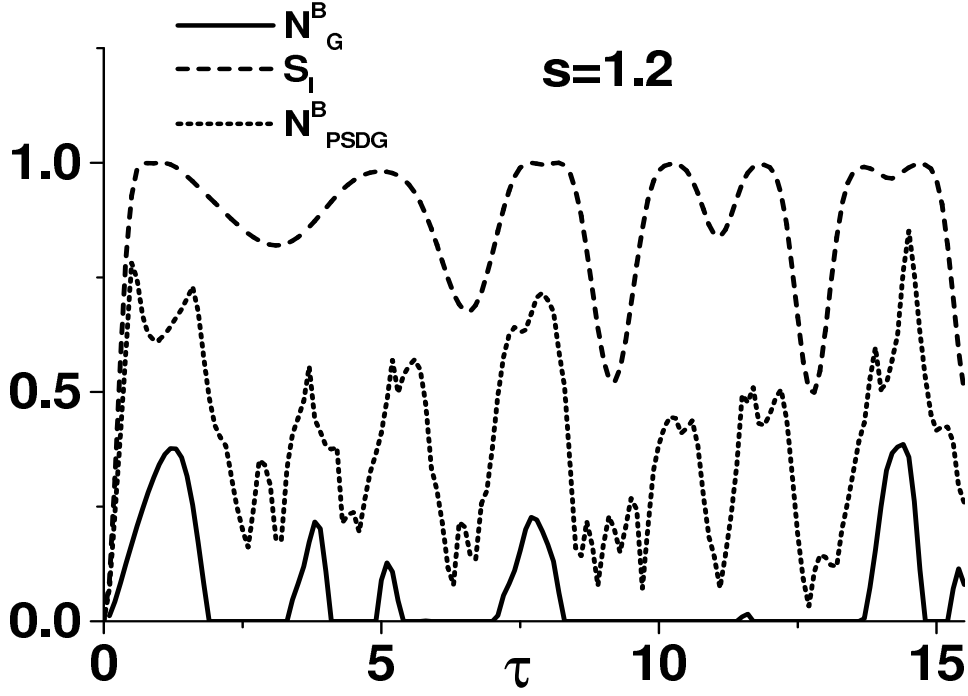}
\caption{The global negativity $N_{G}^{B}$, linear entropy $S_l$, and
negativity $N_{PSDG}^{B}$, versus $\protect\tau (=g\protect\eta t)$ for $%
s=1.2$.}
\label{fig2}
\end{figure}

The global negativity of partial transpose of $\widehat{\rho }%
^{A_{1}A_{2}B}(\tau )$ with respect to $B$ is found to be%
\begin{equation}
N_{G}^{B}=-2\left( \lambda _{1}^{-}+\lambda _{2}^{-}\right) ,
\end{equation}%
where%
\begin{equation*}
\lambda _{1}^{-}=\left( \frac{\rho _{33}^{BA_{1}A_{2}}+\rho
_{55}^{BA_{1}A_{2}}}{2}\right) -\frac{1}{2}\sqrt{\left( \rho
_{33}^{BA_{1}A_{2}}-\rho _{55}^{BA_{1}A_{2}}\right) ^{2}+4\left( \rho
_{26}^{BA_{1}A_{2}}\right) ^{2}},
\end{equation*}%
if 
\begin{equation}
\left( \rho _{33}^{BA_{1}A_{2}}+\rho _{55}^{BA_{1}A_{2}}\right) <\sqrt{%
\left( \rho _{33}^{BA_{1}A_{2}}-\rho _{55}^{BA_{1}A_{2}}\right) ^{2}+4\left(
\rho _{26}^{BA_{1}A_{2}}\right) ^{2}},
\end{equation}%
and zero otherwise. Similarly

\begin{equation*}
\lambda _{2}^{-}=\left( \frac{\rho _{22}^{BA_{1}A_{2}}+\rho
_{44}^{BA_{1}A_{2}}}{2}\right) -\frac{1}{2}\sqrt{\left( \rho
_{22}^{BA_{1}A_{2}}-\rho _{44}^{BA_{1}A_{2}}\right) ^{2}+4\left( \rho
_{15}^{BA_{1}A_{2}}\right) ^{2}},
\end{equation*}%
if 
\begin{equation}
\left( \rho _{22}^{BA_{1}A_{2}}+\rho _{44}^{BA_{1}A_{2}}\right) <\sqrt{%
\left( \rho _{22}^{BA_{1}A_{2}}-\rho _{44}^{BA_{1}A_{2}}\right) ^{2}+4\left(
\rho _{15}^{BA_{1}A_{2}}\right) ^{2}},
\end{equation}%
and $\lambda _{2}^{-}=0$, if%
\begin{equation}
\left( \rho _{22}^{BA_{1}A_{2}}+\rho _{44}^{BA_{1}A_{2}}\right) \geq \sqrt{%
\left( \rho _{22}^{BA_{1}A_{2}}-\rho _{44}^{BA_{1}A_{2}}\right) ^{2}+4\left(
\rho _{15}^{BA_{1}A_{2}}\right) ^{2}}.
\end{equation}%
Furthermore, $E_{3}^{B}=0$, implying that the composite system state does
not have three qubit GHZ state like coherences.

The reduced state operator for qubit $B$ is given by 
\begin{equation}
\widehat{\rho }^{B}(\tau )=Tr_{A_{1}A_{2}}\left( \widehat{\rho }%
^{A_{1}A_{2}B}(\tau )\right) .
\end{equation}%
The purity of a state in Hilbert space of dimension $d$ is measured by the
linear entropy defined as 
\begin{equation}
S_{l}(\rho )=\frac{d}{d-1}\left( 1-tr(\rho ^{2})\right) .
\end{equation}%
The linear entropy is zero for a pure state and one for a maximally mixed
state. Calculated $S_{l}^{B}\left( \widehat{\rho }^{B}(\tau )\right) $ and 
\emph{N}$_{G}^{B}$ are shown as a function of parameter $\tau $ (=$gt$) in
Figure (\ref{fig2}) for $s=1.2$ and $\theta =\pi $. Whereas mixedness of the
state $\widehat{\rho }^{B}(\tau )$ depends on the entanglement of qubit $B$
with atomic qubits and cavity field, the negativity of $\widehat{\rho }%
_{G}^{T_{B}}(\tau )$ is a measure of entanglement of qubit $B$ with qubit
pair $A_{1}A_{2}$. We notice that for certain ranges of parameter $\tau $, $%
\rho _{G}^{T_{B}}$ is found to be positive while $S_{l}^{B}$ value indicates
that the state is entangled. Can we conclude that for $N_{G}^{B}=0$, the
qubit B is entangled only to the cavity field? Since the global negativity
is known to detect only free entanglement of mixed states, the answer is no.
To detect the bound entanglement, we use the pure state decomposition ($PSD$%
) of $\widehat{\rho }^{A_{1}A_{2}B}(\tau )$ given by Eq. (\ref{ropured}),
and calculate 
\begin{equation}
N_{PSDG}^{p}\left( \widehat{\rho }\right) =\sum_{i}p_{i}N_{G}^{p}\left( 
\widehat{\rho }_{i}\right) \text{, }  \label{npsdg}
\end{equation}%
where $p$ stands for one of the three qubits and subscript $PSDG$ refers to
global negativity calculated for a pure state decomposition. In ref. \cite%
{sooj03} the convex roof extension of global negativity ($CRE$) has been
shown to be en entanglement monotone capable of detecting bound
entanglement. The convex roof extension of negativity for a mixed state is
defined as%
\begin{equation}
N_{CRE}^{p}\left( \widehat{\rho }\right) =\min \left(
\sum_{i}p_{i}N_{G}^{p}\left( \widehat{\rho }_{i}\right) \right) ,
\end{equation}%
where $\widehat{\rho }_{i}=\left\vert \Psi _{i}(\tau )\right\rangle
\left\langle \Psi _{i}(\tau )\right\vert $, $\left\vert \Psi
_{i}\right\rangle $ being the eigenvector of $\widehat{\rho }$ corresponding
to eigenvalue $p_{i}$. The negativity $N_{PSDG}^{p}\left( \widehat{\rho }%
\right) $, satisfying the relation 
\begin{equation}
N_{PSDG}^{p}\left( \widehat{\rho }\right) \geq N_{CRE}^{p}\left( \widehat{%
\rho }\right) ,
\end{equation}%
is not an entanglement monotone, but serves to detect bound entanglement.
Numerically calculated $N_{PSDG}^{B}$ for the state $\widehat{\rho }%
^{A_{1}A_{2}B}(\tau )$ is also plotted (dotted line) in Figure (\ref{fig2}).
A comparison with \emph{N}$_{G}^{B}$ plot shows that the regions in which 
\emph{N}$_{G}^{B}=0$, $N_{PSDG}^{B}$ is found to be nonzero, implying that
it is possible to project out entangled states from $\widehat{\rho }%
^{A_{1}A_{2}B}(\tau )$ for any value of $\tau >0$. For \emph{N}$_{G}^{B}=0$,
the state of qubit $B$ is not a separable state in the three qubit mixed
state.

For the state of Eq. (\ref{ropured}), $N_{3}^{p}\left( \widehat{\rho }%
_{i}\right) =0$ for all the pure states $\widehat{\rho }_{i}$ appearing in
the pure state decomposition. This is consistent with the earlier
observation that the state does not have genuine tripartite entanglement.
Genuine tripartite entanglement refers to tripartite entanglement due to
correlations similar to those present in a three qubit GHZ-like state. Loss
of a single qubit destroys this type of entanglement completely, leaving no
residual entanglement. Only two qubit coherences are present in the state $%
\widehat{\rho }^{A_{1}A_{2}B}(\tau )$ as evidenced by $N_{G}^{p}\left( 
\widehat{\rho }_{i}\right) =E_{2}^{p}\left( \widehat{\rho }_{i}\right) $ for
all $\widehat{\rho }_{i}$. As such we may rewrite the Eq. (\ref{npsdg}) as

\begin{equation}
N_{PSDG}^{p}\left( \widehat{\rho }\right) =\sum_{i}p_{i}E_{2}^{p}\left( 
\widehat{\rho }_{i}\right) .  \label{6n}
\end{equation}%
Two qubit coherence can generate not only pairwise entanglement but also
W-like tripartite entanglement. It is common practice to trace out subsystem 
$A_{2}$ to obtain the entanglement of $\ $the pair $A_{1}B$. One can,
however, obtain a measure of $2-$way coherences involving a given pair of
subsystems ($A_{1}B$ or $A_{2}B$ or $A_{1}A_{2}$) from $2-$way partial
transpose constructed from the three qubit state operator $\widehat{\rho }$
by restricting the transposed matrix elements to those for which the state
of the third qubit does not change \cite{shar082}. For example, $\widehat{%
\rho }_{2}^{T_{A_{1}-A_{1}B}}$ is obtained from $\widehat{\rho }$ by
applying the condition%
\begin{eqnarray}
\left\langle i_{1}i_{2}i_{3}\right\vert \widehat{\rho }%
_{2}^{T_{A_{1}-A_{1}B}}\left\vert j_{1}i_{2}j_{3}\right\rangle
&=&\left\langle j_{1}i_{2}i_{3}\right\vert \widehat{\rho }\left\vert
i_{1}i_{2}j_{3}\right\rangle ;\quad if\quad \sum\limits_{m=1}^{3}\left(
1-\delta _{i_{m},j_{m}}\right) =2,\text{ }\quad  \notag \\
\left\langle i_{1}i_{2}i_{3}\right\vert \widehat{\rho }%
_{2}^{T_{A_{1}-A_{1}B}}\left\vert j_{1}j_{2}j_{3}\right\rangle
&=&\left\langle i_{1}i_{2}i_{3}\right\vert \widehat{\rho }\left\vert
j_{1}j_{2}j_{3}\right\rangle ;\quad \text{for all other matrix elements}.
\end{eqnarray}%
We also define the partial transpose $\widehat{\rho }%
_{2}^{T_{A_{1}-A_{1}A_{2}}}$, involving only the pair of subsystems $%
A_{1}A_{2}$, as 
\begin{equation}
\left\langle i_{1}i_{2}i_{3}\right\vert \widehat{\rho }%
_{2}^{T_{A_{1}-A_{1}A_{2}}}\left\vert j_{1}j_{2}i_{3}\right\rangle
=\left\langle j_{1}i_{2}i_{3}\right\vert \widehat{\rho }\left\vert
i_{1}j_{2}i_{3}\right\rangle ,\quad if\quad \sum\limits_{m=1}^{3}\left(
1-\delta _{i_{m},j_{m}}\right) =2,
\end{equation}%
and for all the matrix elements with $i_{3}\neq j_{3}$ 
\begin{equation}
\left\langle i_{1}i_{2}i_{3}\right\vert \widehat{\rho }%
_{2}^{T_{A_{1}-A_{1}A_{2}}}\left\vert j_{1}j_{2}j_{3}\right\rangle
=\left\langle i_{1}i_{2}i_{3}\right\vert \widehat{\rho }\left\vert
j_{1}j_{2}j_{3}\right\rangle .
\end{equation}%
\begin{figure}[t]
\centering \includegraphics[width=3.75in,height=5.0in,angle=-90]{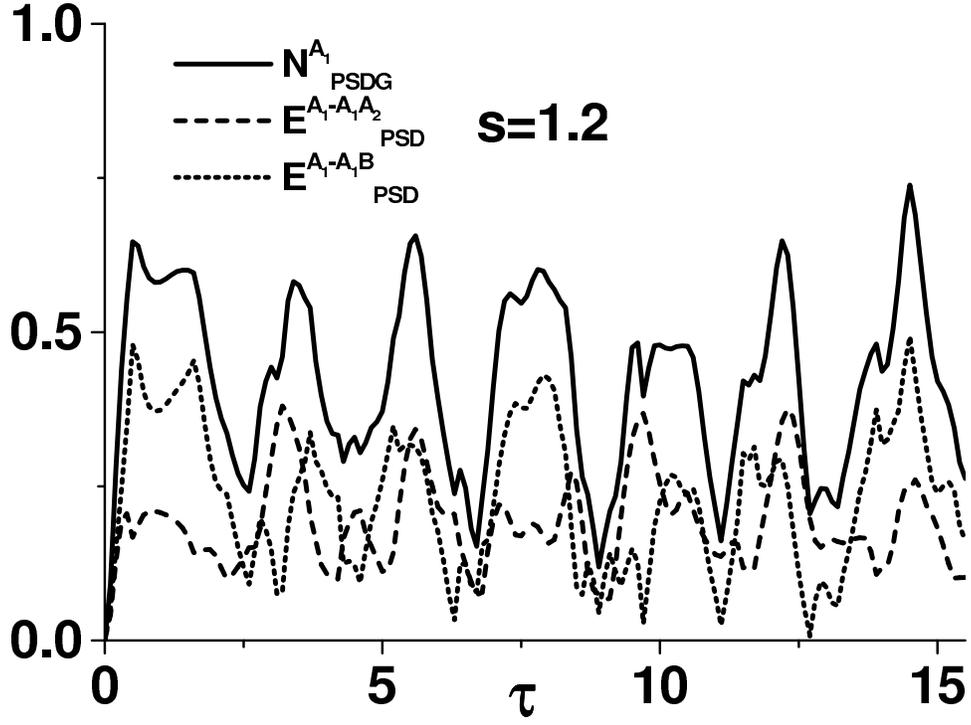}
\caption{The negativity $N_{PSDG}^{A_1}$, $E_{PSD}^{A_1 A_2}$, and $%
E_{PSD}^{A_1 B}$ versus $\protect\tau (=g\protect\eta t)$ for $s=1.2$.}
\label{fig3}
\end{figure}

\begin{figure}[t]
\centering \includegraphics[width=3.75in,height=5.0in,angle=-90]{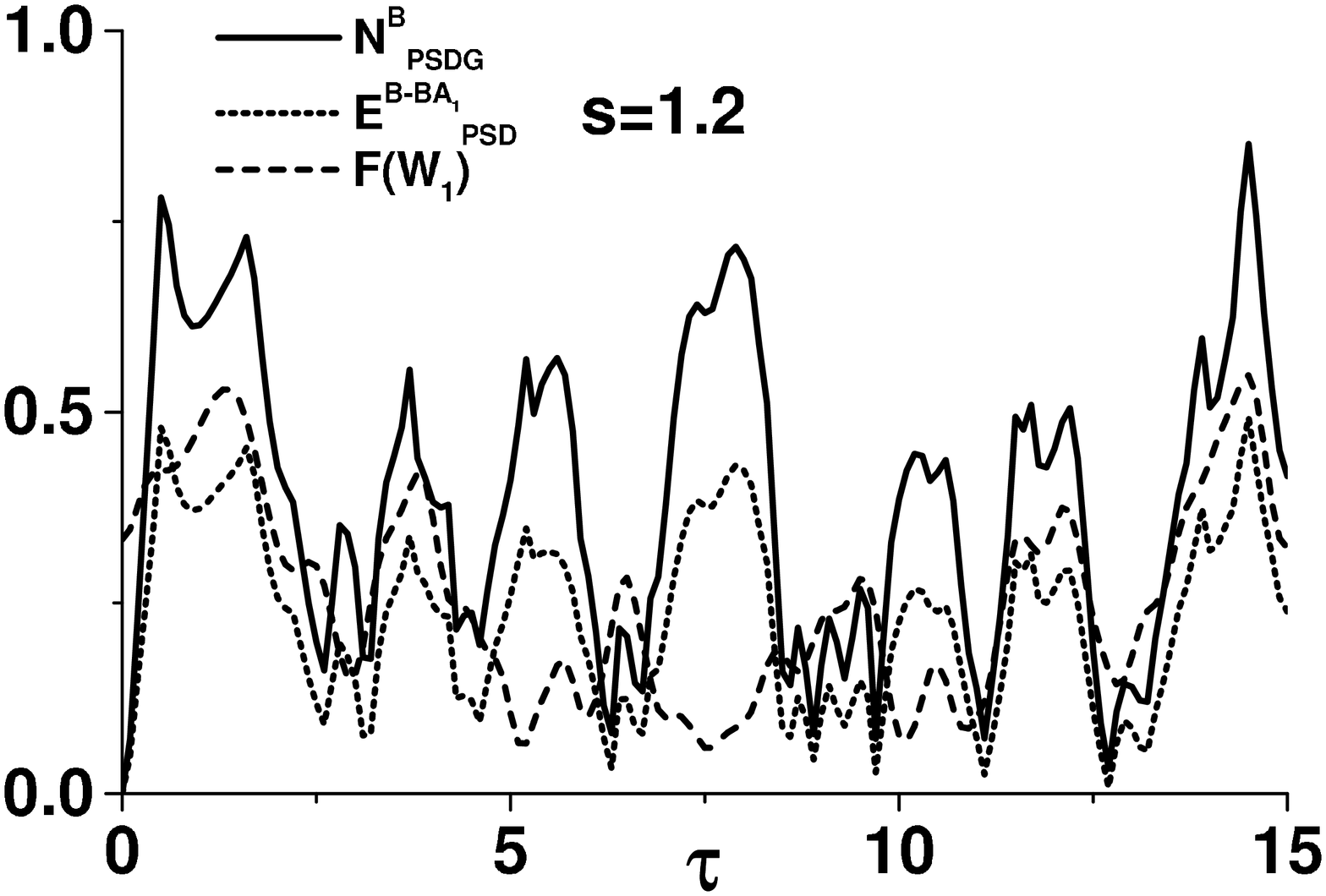}
\caption{The negativity $N_{PSDG}^{B}$, $E_{PSD}^{B A_1 }$ and $F(W_1)$
versus $\protect\tau (=g\protect\eta t)$ for $s=1.2$.}
\label{fig4}
\end{figure}

\begin{figure}[t]
\centering \includegraphics[width=3.75in,height=5.0in,angle=-90]{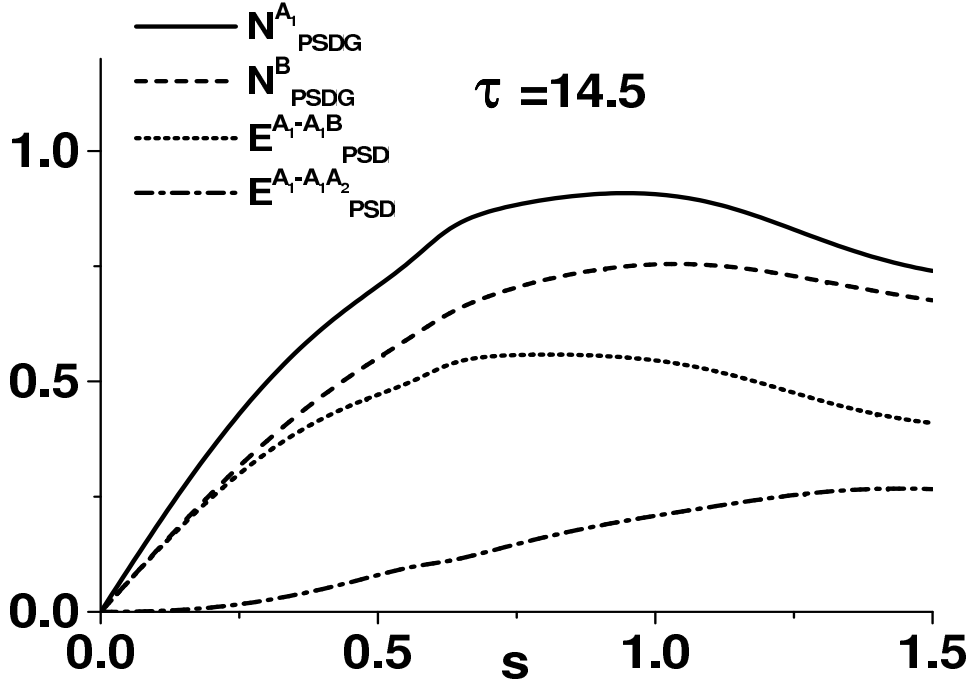}
\caption{The pure state decomposition negativities $N_{PSDG}^{A_{1}}$, $%
N_{PSDG}^{B}$, $E_{PSD}^{A_{1}A_{2}}$, and $E_{PSD}^{A_{1}B}$ versus $s$ for 
$\protect\tau =14.5(=g\protect\eta t)$}
\label{fig5}
\end{figure}
The partial transpose $\widehat{\rho }_{2}^{T_{B-BA_{2}}}$ is defined as 
\begin{equation}
\left\langle i_{1}i_{2}i_{3}\right\vert \widehat{\rho }_{2}^{T_{B-BA_{2}}}%
\left\vert i_{1}j_{2}j_{3}\right\rangle =\left\langle
i_{1}i_{2}j_{3}\right\vert \widehat{\rho }\left\vert
i_{1}j_{2}i_{3}\right\rangle ,\quad if\quad \sum\limits_{m=1}^{3}\left(
1-\delta _{i_{m},j_{m}}\right) =2,
\end{equation}%
and for all the matrix elements with $i_{1}\neq j_{1}$ 
\begin{equation}
\left\langle i_{1}i_{2}i_{3}\right\vert \widehat{\rho }_{2}^{T_{B-BA_{2}}}%
\left\vert j_{1}j_{2}j_{3}\right\rangle =\left\langle
i_{1}i_{2}i_{3}\right\vert \widehat{\rho }\left\vert
j_{1}j_{2}j_{3}\right\rangle \text{.}
\end{equation}%
One can verify that 
\begin{equation}
\widehat{\rho }_{2}^{T_{B}}=\widehat{\rho }_{2}^{T_{B-BA_{1}}}+\widehat{\rho 
}_{2}^{T_{B-BA_{2}}}-\widehat{\rho },  \label{7n}
\end{equation}%
and%
\begin{equation}
\widehat{\rho }_{2}^{T_{A_{1}}}=\widehat{\rho }_{2}^{T_{A_{1}-BA_{1}}}+%
\widehat{\rho }_{2}^{T_{A_{1}-A_{1}A_{2}}}-\widehat{\rho }.  \label{8n}
\end{equation}%
Using the Eqs. (\ref{7n}) and (\ref{8n}) along with the definition of $%
E_{2}^{p}$ in Eq. (\ref{6n}) we get

\begin{eqnarray}
N_{PSDG}^{B}(\widehat{\rho }) &=&E_{PSD}^{B-BA_{1}}(\widehat{\rho }%
)+E_{PSD}^{B-BA_{2}}(\widehat{\rho }), \\
N_{PSDG}^{A_{1}}(\widehat{\rho }) &=&E_{PSD}^{A_{1}-A_{1}A_{2}}(\widehat{%
\rho })+E_{PSD}^{A_{1}-A_{1}B}(\widehat{\rho }),
\end{eqnarray}%
where partial negativity $E_{PSD}^{B-BA_{1}}(\widehat{\rho })$ or $%
E_{PSD}^{B-BA_{2}}(\widehat{\rho })$, associated with pure state
decomposition, is defined as 
\begin{equation}
E_{PSD}^{B-BA}\left( \widehat{\rho }\right) =\sum_{i}p_{i}\sum_{\mu
}\left\langle \Phi _{\mu ,i}^{-}\right\vert \left( \widehat{\rho }%
_{i}\right) _{2}^{T_{B-BA}}\left\vert \Phi _{\mu ,i}^{-}\right\rangle ,
\end{equation}%
and $\Phi _{\mu ,i}^{-}$ is the $\mu ^{th}$ eigenvector corresponding to
negative eigenvalue $\lambda _{\mu ,i}$ of the partial transpose $\left( 
\widehat{\rho }_{i}\right) _{2}^{T_{B}}$. The partial negativities $%
E_{PSD}^{A_{1}-A_{1}A_{2}}(\widehat{\rho })$ and $E_{PSD}^{A_{1}-A_{1}B}(%
\widehat{\rho })$ are given by%
\begin{eqnarray*}
E_{PSD}^{A_{1}-A_{1}A_{2}}\left( \widehat{\rho }\right)
&=&\sum_{i}p_{i}\sum_{\mu }\left\langle \chi _{\mu ,i}^{-}\right\vert \left( 
\widehat{\rho }_{i}\right) _{2}^{T_{A_{1}-A_{1}A_{2}}}\left\vert \chi _{\mu
,i}^{-}\right\rangle , \\
E_{PSD}^{A_{1}-A_{1}B}\left( \widehat{\rho }\right)
&=&\sum_{i}p_{i}\sum_{\mu }\left\langle \chi _{\mu ,i}^{-}\right\vert \left( 
\widehat{\rho }_{i}\right) _{2}^{T_{A_{1}-A_{1}B}}\left\vert \chi _{\mu
,i}^{-}\right\rangle ,
\end{eqnarray*}%
where $\chi _{\mu ,i}^{-}$ is the $\mu ^{th}$ eigenvector corresponding to
negative eigenvalue $\beta _{\mu ,i}$ of the partial transpose $\left( 
\widehat{\rho }_{i}\right) _{2}^{T_{A_{1}}}$.

The partial $K-$way negativities $E_{PSD}^{B-BA_{1}}(\widehat{\rho })$, $%
E_{PSD}^{B-BA_{2}}(\widehat{\rho })$ and $E_{2}^{A_{1}-A_{1}A_{2}}$
calculated from pure state decomposition of state operator determine the
pairwise entanglement of qubit pairs $BA_{1}$, $BA_{2}$, and $A_{1}A_{2}$.
The entanglement available to $A_{1}$ and $B$ if they have no knowledge of $%
A_{2}$, as well as the probabilistic entanglement of subsystem $BA_{1}$
after a measurement has been made by $A_{2}$, is determined by $%
E_{PSD}^{B-BA_{1}}(\widehat{\rho })$. The eigenvalues and eigenvectors of $%
\widehat{\rho }^{A_{1}A_{2}B}(\tau )$ are used to calculate, numerically,
the partial negativities $E_{PSD}^{A_{1}-A_{1}A_{2}}\left( \widehat{\rho }%
\right) $, $E_{PSD}^{A_{1}-A_{1}B}\left( \widehat{\rho }\right) $ and \ $%
E_{PSD}^{B-BA_{1}}\left( \widehat{\rho }\right) =E_{PSD}^{B-BA_{2}}\left( 
\widehat{\rho }\right) $. Figure (\ref{fig3}) displays $N_{PSDG}^{A_{1}}%
\left( \widehat{\rho }\right) =N_{PSDG}^{A_{2}}\left( \widehat{\rho }\right) 
$, $E_{PSD}^{A_{1}-A_{1}A_{2}}\left( \widehat{\rho }\right) $, and $%
E_{PSD}^{A_{1}-A_{1}B}\left( \widehat{\rho }\right) $ \ versus parameter $%
\tau $ for $s=1.2$ and $\theta =\pi $. The negativity $N_{PSDG}^{B}\left( 
\widehat{\rho }\right) $, and $E_{PSD}^{B-BA_{1}}\left( \widehat{\rho }%
\right) =E_{PSD}^{B-BA_{2}}\left( \widehat{\rho }\right) $ are plotted in
figure (\ref{fig4}) along with the fidelity $F(W_{1})$ defined as in Eq. (%
\ref{fidelity}). The maxima in $F(W_{1})$ plot are located in the regions
with large $E_{PSD}^{B-BA_{1}}\left( \widehat{\rho }\right) $ and finite $%
E_{PSD}^{A_{1}-A_{1}A_{2}}\left( \widehat{\rho }\right) $, consistent with
the fact that the reduced two qubit states $\widehat{\rho }^{BA_{1}}(\tau ),$
$\widehat{\rho }^{BA_{2}}(\tau ),$ and $\widehat{\rho }^{A_{1}A_{2}}(\tau )$
are entangled states. The partial negativities $E_{PSD}^{B-BA_{1}}\left( 
\widehat{\rho }\right) $, $E_{PSD}^{B-BA_{2}}\left( \widehat{\rho }\right) $
and $E_{PSD}^{A_{1}-A_{1}A_{2}}\left( \widehat{\rho }\right) $ detect the
pairwise entanglement present in the composite state. Simultaneously nonzero
positive valued $E_{PSD}^{A_{1}-A_{1}A_{2}},$ $E_{PSD}^{A_{1}-A_{1}B}$, and $%
E_{PSD}^{B-BA_{2}}\left( \widehat{\rho }\right) $, signal the W-like
tripartite entanglement of three qubit state $\widehat{\rho }%
^{A_{1}A_{2}B}(\tau )$.

The calculated negativities $N_{PSDG}^{A_{1}}\left( \widehat{\rho }\right) $%
, $N_{PSDG}^{B}\left( \widehat{\rho }\right) $, $E_{PSD}^{A_{1}-A_{1}B}%
\left( \widehat{\rho }\right) $ and $E_{PSD}^{A_{1}-A_{1}A_{2}}\left( 
\widehat{\rho }\right) $ versus squeeze parameter $s$ at $\tau =14.5$ for $%
\theta =\pi $, are shown in Fig. (\ref{fig5}). The decrease in the
probability of finding the cavities in vacuum state with increase in the
value of $s$ results in an increase in the probability of finding the system
in an entangled state. The peak value of $N_{PSDG}^{B}\left( \widehat{\rho }%
\right) $ occurs for $s=0.95$ while $N_{PSDG}^{A_{1}}\left( \widehat{\rho }%
\right) $ is maximized at $s=1.04.$ We have selected $s=1.2$ as the squeeze
parameter for the reason that the probability of finding the qubits $A_{1}$,
and $A_{2}$ in entangled state is appreciable for this choice.

\section{Conclusions}

We have studied the entanglement generation through interaction of squeezed
light, shared by two remote cavities, with two atoms in cavity $c_{1}$ and a
single atom in cavity $c_{2}$. No direct interaction amongst the atoms takes
place. Analytical expressions for the matrix elements of the three atom
mixed state as a function of atom-field coupling strength and interaction
time, are obtained. The decrease in the probability of finding the cavities
in vacuum state with increase in the value of squeeze parameter $s$ results
in an increase in the probability of finding the system in an entangled
state. Numerical calculations to study the dynamics of three atom
entanglement show the qubit pair $A_{1}A_{2}$ in Bell-like state entangled
to remote qubit $B$. From the reduced state operators for three atom system,
the linear entropy is calculated and compared with negativities of partially
transposed state operators as well as negativities calculated from the pure
state decomposition of the state operator. The bound entanglement of the
mixed state is not detected by the global negativity, however, the
negativity $N_{PSDG}^{B}\left( \widehat{\rho }\right) $ calculated from the
pure state decomposition of the state operator detects the bound
entanglement. The degree of entanglement depends on the squeeze parameter as
well as the coefficient of reflection of the beam splitter. For $\tau =14.5$%
, the pure state decomposition negativity is seen to be maximum for qubit $%
A_{1}$ when $s=1.04$, whereas the peak value of $N_{PSDG}^{B}\left( \widehat{%
\rho }\right) $ occurs for $s=0.95.$ A higher value of squeeze parameter, $%
s=1.2,$ has been selected to study the entanglement dynamics of the system
since the probability of finding the qubits $A_{1}$ and $A_{2}$ in entangled
state is higher than that for $s=1.04$. The partial negativities calculated
by selective partial transposition of the three atom mixed state detect the
pairwise entanglement of qubit pairs $A_{1}B$, $A_{2}B$, and $A_{1}A_{2}$.
The entanglement of three atoms is found to be W-like, no three qubit GHZ
like quantum correlations being generated. It is well known, however, that a
GHZ state can be distilled from a W-like tripartite state, by stochastic
local operations and classical communication. Trapped atoms with long lived
metastable states are favored as memory qubits. The advantage of having
qubits $A_{1}$ and $A_{2}$ in Bell-like entangled state is that the
entanglement between qubits $A_{1}$ and $B$ can be enhanced through local
operations on qubit $A_{2}$. The entangled states, generated in the scheme
proposed here, can be used for implementing quantum communication protocols
and for quantum computation.

\subsection{Acknowledgements}

Financial support from Capes, Fund\c{c}\~{a}o araucaria and FAEP-UEL, Brazil
is acknowledged.

\end{document}